\documentclass[superscriptaddress,twocolumn,secnumarabic,
 amssymb,amsmath,nobibnotes,aps,prd,showkeys,showpacs]{revtex4}

\usepackage{graphicx}
\usepackage{bm}%
 
\begin{document}


\title{Dynamics and constraints of the Unified Dark Matter flat cosmologies}

\author{Spyros Basilakos}
\affiliation{Academy of Athens, Research Center for Astronomy and Applied Mathematics,
 Soranou Efesiou 4, GR-11527, Athens, GREECE}

\author{Georgios Lukes-Gerakopoulos}
\affiliation{Academy of Athens, Research Center for Astronomy and Applied Mathematics,
 Soranou Efesiou 4, GR-11527, Athens, GREECE}
\affiliation{University of Athens,Department of Physics, Section of Astrophysics,
 Astronomy and Mechanics}

\begin{abstract}
We study the dynamics of the scalar field FLRW flat cosmological models 
within the framework of the {\it Unified Dark Matter} (UDM) scenario. In this model we find 
that the main cosmological functions such as the scale factor of the Universe, the scalar
field, the Hubble flow and the equation of state parameter are defined in 
terms of hyperbolic functions. These analytical solutions  
can accommodate an accelerated
expansion, equivalent to either the {\em dark energy} or the standard $\Lambda$ models.
Performing a joint likelihood analysis 
of the recent supernovae type Ia data  
and the Baryonic Acoustic Oscillations traced by the Sloan Digital 
Sky Survey (SDSS) galaxies, 
we place tight constraints on the main cosmological parameters 
of the UDM cosmological scenario. 
Finally, we compare the UDM scenario with various 
dark energy models namely $\Lambda$ cosmology,
parametric dark energy model and variable Chaplygin gas. 
We find that the UDM scalar field model provides a large and 
small scale dynamics which are in fair agreement with the 
predictions by the above dark energy models 
although there are some differences especially at high redshifts.
\end{abstract}
\maketitle

\section{Introduction}
The detailed analysis of the available high quality cosmological 
data (Type Ia supernovae \cite{Riess07}, \cite{essence}; 
CMB \cite{Spergel07}, \cite{Komatsu08}, etc.) 
leads to the conclusion that we live in a flat 
and accelerating universe. 
In order to investigate the cosmic history
of the observed universe, we have to introduce a general cosmological model which 
contains cold dark matter to explain the large scale structure clustering and an extra 
component with negative pressure, the vacuum energy (or in a more general setting
the ``dark energy''), to explain the observed accelerated cosmic expansion 
(Refs. \cite{Riess07,essence,Spergel07,Komatsu08} and references therein). 
The nature of the dark energy is one of the most fundamental and 
difficult problems in physics and cosmology. 
There are many theoretical speculations regarding the physics of 
the above exotic dark energy, 
such as a cosmological constant (vacuum), quintessence, $k-$essence, vector fields, 
phantom, tachyons, Chaplygin gas and the list goes on
(see \cite{Ratra88,Weinberg89,Wetterich:1994bg,Caldwell98,KAM,Caldwell,Peebles03,
Brax:1999gp,fein02,chime04,Brookfield:2005td,Copel06,Boehmer:2007qa,Friem08} 
and references therein).

Such studies are based on the general assumption that
the real scalar field $\phi$ rolls down the potential $V(\phi)$ 
and therefore it  
could resemble the dark energy 
\cite{Ozer87,Peebles88,
Weinberg89,Turner97,Caldwell98,Padm03,Peebles03}. 
This is very important because scalar fields could provide possible solutions 
to the cosmological coincidence problem. 
In this framework, 
the corresponding stress-energy tensor takes the form of a perfect fluid, with
density $\rho_{\phi}={\dot \phi}^{2}/2+V(\phi)$ and pressure $P_{\phi}={\dot \phi}^{2}/2-V(\phi)$. 
From a cosmological point of view, if the scalar field varies slowly
with time, so that ${\dot \phi}^{2}/2V \ll 1$, then ${\rm w}\equiv P_{\phi}/\rho_{\phi}\approx -1$,
which means that the scalar field evolves like a vacuum energy. 
Of course in order to investigate the overall dynamics 
we need to define the functional form of the 
potential energy. 
The simplest example found in the literature is a scalar field 
with $V(\phi)\propto \phi^{2}$ 
(see for review \cite{Peebles03}, \cite{Dolgov90}) 
and it has been shown that the time evolution of this scalar field is dominated 
by oscillations around $\phi=0$. 
Of course, the issue of the potential energy has a long history in scalar field cosmology 
(see \cite{sahni00,santi00,sen02,kehagias04,Gorini05} and
references therein) 
and indeed several parameterizations have been proposed  
(exponential, power law, hyperbolic etc).

The aim of the present work is to investigate the observational consequences 
of the overall dynamics of a family of flat cosmological models
by using a hyperbolic scalar field potential which appears to 
act both as dark matter and dark energy 
\cite{Bertacca07}. To do so, we use the traditional Hamiltonian approach.
In fact, the idea to build cosmological models in which 
the dark energy component is somehow linked with the 
dark matter is not new in this kind of studies.
Recently, alternative approaches to the unification 
of dark energy and dark matter have been proposed in 
the framework of the generalized Chaplygin gas \cite{Kam01,Bili02} 
and in the context of supersymmetry \cite{Taka06}.

The structure of the paper is as follows. 
The basic theoretical elements of the problem are 
presented in section 2 by solving analytically [for spatially flat 
Unified Dark Matter (UDM) scalar field models]
the equations of motion. In section 3, 
we present the functional forms of the 
basic cosmological functions [$a(t)$, $\phi(t)$ and $H(t)$].
In section 4 we place constraints on the main parameters of our model by
performing a joint likelihood analysis utilizing the SNIa data 
\cite{essence} and the observed Baryonic Acoustic Oscillations (BAO) 
\cite{Eis05} and \cite{Pad07}. 
In particular, we find that the matter density at the present time 
is $\Omega_{m} \simeq 0.25$ while the corresponding  
scalar field is  
$\phi_{0} \simeq 0.42$ in geometrical units (0.084 in Planck units).  
Section 5 outlines the evolution of matter
perturbations in the UDM model. Also we compare the theoretical
predictions provided by the UDM scenario with those 
found by three different type of dark energy models 
namely $\Lambda$ cosmology,
parametric dark energy model and variable Chaplygin gas. 
We verify that at late times (after the inflection point) 
the dynamics of the UDM scalar model 
is in a good agreement, 
with those predicted by the above dark energy models
although there are some differences especially at early epochs:
(i) the UDM equation of state 
parameter takes positive values at large redshifts, (ii) 
it behaves well with respect to the cosmic coincidence problem, 
and (iii) before the inflection point 
the cosmic expansion in the UDM model 
is much more decelerated than in the other three
dark energy models implies that the large scale structures 
(such as galaxy clusters) are more
bound systems with respect to those cosmic structures which 
produced by the other three dark energy models.
Finally, we draw our conclusions in section 6.

\section{Analytical solutions in the flat scalar field cosmology}
Within the framework of 
homogeneous and isotropic scalar field cosmologies
it can be proved (see Ref. \cite{Page84})
that the main cosmological equations 
(the so called Friedmann-Lemaitre equations) can  
be obtained by a Lagrangian formulation:
\begin{equation}
      L=-3a \dot{a}^2 + a^3 \left[\frac{\dot{\phi}^2}{2}-V(\phi)\right]+3ka 
      \label{Langaf}
\end{equation}
where $a(t)$ is the scale factor of the universe, 
$\phi(t)$ is the scalar field, $V(\phi)$ is the potential energy 
and $k(=-1,0,1)$ is the spatial curvature. Indeed the equations of motion 
take the following forms\footnote{Note that in this work we set $8 \pi G=c\equiv 1$
which corresponds to ${\cal D}^{2}=3/8$. For Planck units we have to set 
$G=c\equiv 1$ with ${\cal D}^{2}=3\pi $. For S.I units we have 
${\cal D}^{2}=3\pi G/c^{2}$.}:

\begin{eqnarray}
     3 \left[\left( \frac {\dot{a}} {a} \right )^2+\frac{k}{a^{2}}\right]&=&
     \frac{\dot{\phi}^2}{2}+V(\phi) \label{EFE1} \\
     2 \left(\frac{\ddot{a}}{a}\right)+\left(\frac {\dot{a}} {a} 
     \right)^2+\frac{k}{a^{2}}&=&-\frac{\dot{\phi}^2}{2}+V(\phi)
     \label{EFE2}
\end{eqnarray}
and 
\begin{equation}
      \ddot{\phi}+3 \frac{\dot{a}}{a}\dot{\phi}+V^{\prime}(\phi)=0\;\;
      \label{SEC}
\end{equation}
where the over-dot denotes derivatives with respect to time while  
prime denotes derivatives with respect to $\phi$. 
We would like to stress here that 
in this work we consider a 
spatially homogeneous scalar field $\phi$, ignoring the possible coupling
to other fields and quantum-mechanical effects. 
On the other hand, introducing in the global dynamics a new 
degree of freedom, in a form of the scalar field $\phi$, 
it is possible to make the vacuum 
energy a function of time (see \cite{Peebles88}, \cite{wetter}, \cite{zlatev}).    
Note of course that the geometry of the space-time is described by the 
Friedmann-Lemaitre-Robertson-Walker (FLRW) line element.

In order to study the above system of differential equations we need to 
define explicitly the functional form of the scalar field potential
energy, $V(\phi)$, which is not an easy task to do. 
Indeed, in the literature, due to the unknown nature of the dark energy,
there are many forms of potentials proposed by several authors 
(for a review see \cite{Caldwell}, \cite{Copel06}, 
 \cite{Toporensky06}) 
which describe differently the physics of the scalar field. 
It is worth pointing out that for some special cases
analytical solutions have been found
(Refs. \cite{Turner83,mata85,stein99,santi00,sen02,kehagias04,Gorini05} and
references therein). As an example,
if the potential $V(\phi)$ is modeled as a 
power law $\phi^{n}$, then the energy 
density of the scalar field evolves like $\rho_{\phi} \propto
a^{-6n/(n+2)}$ which means that, for $n=2$ or $n=4$ 
the corresponding energy density behaves either like non relativistic or relativistic matter.
In this work, we have used a functional form of $V(\phi)$ (see \cite{Lukes08}) 
for which we solve the previous dynamical problem analytically. This potential 
corresponds to the so called {\it Unified Dark Matter} (hereafter UDM) 
scenario \cite{Gorini05}, \cite{Bertacca07}, \cite{Gorini04}:
\begin{equation}
      \label{potent} 
      V(\phi) =c_{1} \cosh^2{({\cal D}~\phi)}+c_2   ~~~~~  {\cal D}, c_1,~ c_2 ~ \in ~ \Re \;\;.
\end{equation}

Following the Bertacca et al. \cite{Bertacca07} nomenclature, 
the real constants in eq. (\ref{potent}) 
are selected such as $ c_1=c_2>0 $.  
As expected there is one minimum at the point $\phi=0$, which reads 
\begin{equation}
V_{min}=V(0)=c_1+c_2 \;\;. 
\end{equation}
We would like to point out that 
as long as the scalar field is taking negative and large values 
the UDM model has the attractive 
feature due to $V(\phi) \propto e^{-2{\cal D}\phi}$ \cite{sahni00}.
This property simply says that the energy density in $\phi$ tracks \cite{stein99}
the radiation (matter) component.
In fact the UDM potential was designed to 
mimic both the dark matter and the 
dark energy. Indeed, performing a Taylor expansion to the 
potential around its minimum we get,
\begin{equation}
V(\phi)=V_{min}+c_{1} {\cal D}^{2} \phi^{2}+\frac{c_{1} {\cal D}^{4} }{3}\phi^{4}+.. 
\end{equation}
which means that at an early enough epoch the ''cosmic'' fluid behaves 
like radiation \cite{Scherrer04} ($V(\phi) \propto \phi^{4}$), then evolves to 
the matter epoch ($V(\phi) \propto \phi^{2}$) and 
finishes with a phase that looks like a cosmological 
constant (see also \cite{Bertacca07}).

Changing now the variables from $(a,\phi)$ to $(x_{1},x_{2})$ using the relations:
\begin{eqnarray} 
      \label{trans}
      x_{1}=\sqrt{\frac{8}{3}} ~a^{3/2} \sinh{({\cal D} ~\phi)} \nonumber \\
      x_{2}=\sqrt{\frac{8}{3}} ~a^{3/2} \cosh{({\cal D} ~\phi)}
\end{eqnarray}
with ${\cal D}^{2} =3/8$ the Lagrangian (\ref{Langaf}) is written:
\begin{eqnarray}
      \label{Langxy}
      L=\frac{1}{2}\left[(\dot{x_{1}}^2+\frac{3}{4}c_2~ x^2_{1})-
       (\dot{x_{2}}^2+\frac{3}{4}(c_1+c_2)~ x^2_{2})\right] \nonumber \\ 
        +\frac{1}{2}\left[3^{4/3} k (x_2^2-x_1^2)^{1/3}\right] \;\;.
\end{eqnarray}
The scale factor ($a>0$) in the UDM scenario is now given by:
\begin{equation}
      a = \left[ \frac{3(x_{2}^2-x_{1}^2)}{8}\right] ^{1/3} \;\;,
      \label{alcon}
\end{equation}
which means that the new variables have to satisfy the following inequality:
$x_{2} \ge |x_{1}|$.

It is straightforward now from the Lagrangian 
(\ref{Langxy}) to write the corresponding Hamiltonian:
\begin{eqnarray}
      \label{Hamxy1}
  {\cal H} =\frac{1}{2}\left[(p_{x_{1}}^2-\omega_1^2 x^2_{1})-
     (p_{x_{2}}^2-\omega_2^2 x^2_{2})\right] \nonumber \\
    -\frac{1}{2}\left[3^{4/3} k (x_2^2-x_1^2)^{1/3}\right] \;\;. 
\end{eqnarray}
where $ p_{x_{1}}=\dot{x_{1}} $, $ p_{x_{2}}=-\dot{x_{2}}$ denote the canonical momenta and 
$ \omega_1^2 = \frac{3}{4}c_2 $, $ \omega_2^2 = \frac{3}{4}(c_1+c_2) $
are the oscillators' ``frequencies'' 
with units of inverse of time and 
\begin{eqnarray}
      \label{om11}
\frac{\omega_{2}^{2}}{\omega_{1}^{2}}=1+\frac{c_{2}}{c_{1}}=\kappa \;\;\;\;\;(\kappa=2)\;\;\;.
\end{eqnarray}

The dynamics of the closed FRLW scalar field cosmologies has been 
investigated thoroughly in \cite{Lukes08}. 
In particular, for a semi-flat
geometry ($k\longrightarrow 0$) we have revealed cases where the dynamics of the system
(see section 3.1 in \cite{Lukes08}, orbit 5 in Fig. 1 
scale factor vs. time and Fig.4) is close to the 
concordance $\Lambda$-cosmology, despite the fact
that for the semi flat UDM model there is a strong indication of a chaotic
behavior. In this paper we would like to investigate the potential 
of a spatially flat UDM scenario ($k=0$) since the analysis of the Cosmic 
Microwave Background (CMB) anisotropies have strongly suggested that the 
spatial geometry of the universe is flat \cite{Spergel07}. 
Technically speaking, in the new coordinate system our dynamical problem is 
described well by two independent hyperbolic oscillators and thus the system is fully 
integrable. Indeed in the new coordinate system the corresponding equations of motion
can be written as
\begin{eqnarray}
      \label{eqmot1}
      \dot{p}_{x_{1}}=-\frac{\partial{\cal H}}{\partial{x_{1}}} = \omega_1^2x_{1} &,&
      \dot{p}_{x_{2}}=-\frac{\partial{\cal H}}{\partial{x_{2}}} = -\omega_2^2x_{2} 
      \nonumber 
\end{eqnarray}
and it is routine to perform the integration to find the analytical solutions:
\begin{eqnarray} 
      \label{trans11}
      x_{1}(t)=A_1 \sinh{(\omega_{1}t+\theta_{1})} \nonumber \\
      x_{2}(t)=A_2 \sinh{(\omega_{2}t+\theta_{2})} 
\end{eqnarray}
where $A_1$, $A_2$, $\theta_{1}$ and $\theta_2$ are the 
integration constants of the problem. 
With the aid of eq.(\ref{trans11}) and
assuming that the total energy ($ {\cal H} $) of the system is zero
the above constants satisfy the following restriction: 
\begin{eqnarray} 
A_{1}^{2}\omega_{1}^{2}-A_{2}^{2}\omega_{2}^{2}=0
\Rightarrow \frac{A_{1}^{2}}{A_{2}^{2}}=\frac{\omega_{2}^{2}}{\omega_{1}^{2}}=\kappa \;\;.
\end{eqnarray} 
As expected the phase space of the current dynamical problem 
is simply described by two hyperbolas 
$p_{x_{i}}^{2}-\omega_{i}^{2}x_{i}^{2}=\omega_{i}^{2}A_{i}^{2}$
whose axes have a ratio $1/\omega_{i}$ ($i=1,2$). 

 \begin{figure}
         \centerline{\includegraphics[width=20pc] {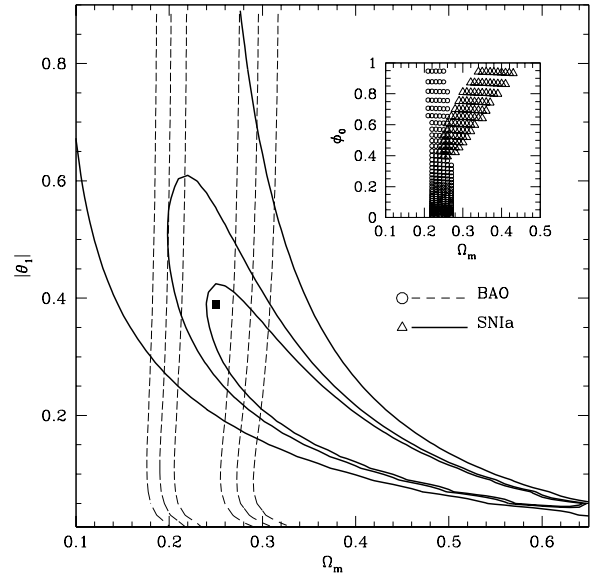}} 
  \caption{Likelihood contours in the $(\Omega_{m},|\theta_{1}|)$ plane.
The contours correspond to 1$\sigma$, 2$\sigma$  and 3$\sigma$
confidence levels. The thick (thin) contours correspond to the 
SNIa (BAOs) likelihoods while the solid square is the best fit solution: 
$\Omega_{m}\simeq 0.25$ and $\theta_{1}\simeq -0.39$. {\it Insert Panel:}
The solutions within $1\sigma$ contours in the $(\Omega_{m},\phi_{0})$ plane,
where $\phi_{0}$ is the present value of the scalar field in geometrical 
units ($8\pi G=c\equiv 1$). Note that
for Planck units ( $G=c\equiv 1$) we have to multiply the $\phi_{0}$-axis with
$(8 \pi)^{-1/2}$. Using the 
best fit solution we find $\phi_{0}\simeq 0.42$ or $0.084$ in Planck units.}
         \label{Figcon}
 \end{figure}

\section{The Evolution of the UDM cosmological functions}
In this section, with the aid of the basic hyperbolic functions, we 
analytically derive
the predicted time dependence of the main cosmological functions in
the UDM cosmological model.

\subsection{Scalar field - potential versus time}
If we combine eq.(\ref{trans11}) together 
with the initial parameterization [see eq.(\ref{trans})]
we immediately obtain the following expressions:
\begin{eqnarray} 
      \label{psi11} 
\frac{x_{1}}{x_{2}}=\tanh({\cal D}\phi)=\frac{\sqrt{\kappa}\sinh(\omega_1t+\theta_{1})}{
\sinh(\omega_{2}t+\theta_{2})}=\Psi(t)
\end{eqnarray} 
and after some algebra the evolution of the scalar field becomes:
\begin{eqnarray} 
      \label{phi11} 
\phi(t)=\frac{1}{2{\cal D}}{\rm ln}\left[\frac{1+\Psi(t)}{1-\Psi(t)}\right] \;\;.
\end{eqnarray}
Using eqs. (\ref{potent}) and (\ref{phi11}) one can prove that  
\begin{eqnarray} 
      \label{potent11} 
V(t)=\frac{4\omega^{2}_{1}}{3}\left[ \frac{\kappa-\Psi^{2}(t)}{1-\Psi^{2}(t)}\right] \;\;.
\end{eqnarray}
Now the range of $\Psi$-values for which the UDM scalar field is well defined [due to 
eq.(\ref{phi11})] is: $\Psi \in (-1, 1)$. 
Evidently, when the system reaches at the critical point $\phi=0$ then 
$\Psi(t_{m})=0$ (or $V_{min}=4\omega^{2}_{1}\kappa/3$). For this to be the case
we must have $t_{m}=-\theta_{1}/\omega_{1}$ and therefore, $\theta_{1}<0$.

 \begin{figure}
         \centerline{\includegraphics[width=20pc] {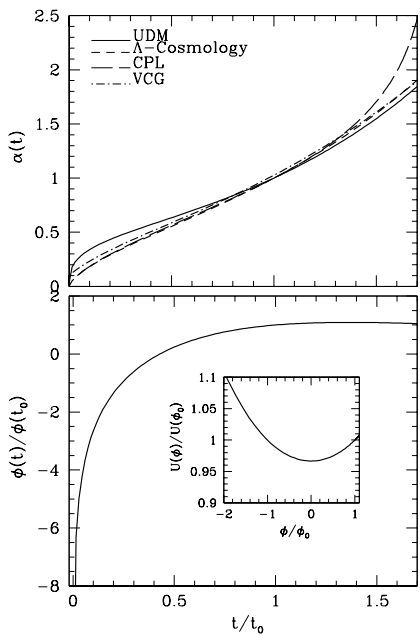}} 
  \caption{
{\it Upper Panel:} Comparison of the scale factor provided by the 
UDM model (solid line) with the 
traditional $\Lambda$ cosmology (short dashed line), VCG (dot dashed)
and CPL (long dashed) dark energy models.
{\it Bottom Panel:} The evolution of the scalar field.  
In the insert panel we present the behavior of the potential 
normalized to unity at the present time. Note, that 
$t_{0}\approx H^{-1}_{0}\simeq 13.6$Gyr is the present age of the universe.
  }
         \label{Fighubmod}
 \end{figure}

\subsection{Scale factor - Hubble flow versus time} 
Inserting eq.(\ref{trans11}) into eq.(\ref{alcon}) the scale factor, 
normalized to unity at the present epoch, evolves in time as 
\begin{eqnarray}
      \label{a11} 
a(t)\equiv \frac{a}{a_{0}}=\left[\frac{
\sinh^{2}{(\omega_{2}t+\theta_{2})}-\kappa\sinh^{2}{(\omega_{1}t+\theta_{1})}}
{\sinh^{2}{(\omega_{2}t_{0}+\theta_{2})}-\kappa\sinh^{2}{(\omega_{1}t_{0}+\theta_{1})}}
\right]^{1/3}
\end{eqnarray} 
where $t_{0}$ is the present age of the universe in billion years.
The constant $\theta_{1}$ is related to $\theta_{2}$ because
at the singularity ($t=0$), the scale factor has to 
be exactly zero [see eq.(\ref{a11})]. 
After some algebra, we find that
\begin{eqnarray}
\label{con2} 
\theta_{2}={\rm ln}\left(-\sqrt{\kappa}\sinh\theta_{1}+\sqrt{\kappa\sinh^{2}\theta_{1}+1} 
\right)\ge 0 \;\;.
\end{eqnarray}

Furthermore, we investigate the circumstances
under which an inflection point exists 
and therefore have an acceleration phase of the scale factor. 
This crucial period in the cosmic history corresponds to 
$\ddot{a}(t_{I})=0$ which implies that the condition 
$$V(\phi_{I})-\dot{\phi_{I}}^{2}=0$$ 
should contain roots
which are real and such that $a \in (0,1)$. The above equation
is solvable because for $c_1, c_2>0$ the potential energy [$V(\phi)$] 
takes only positive values. Knowing the integration constants 
($\omega_{1},\theta_{1},\omega_{2},\theta_{2}$), of the current dynamical problem we can 
calculate the inflection point
by solving numerically the following equation:
\begin{eqnarray} 
2\dot{\Psi}^{2}(t_{I})-\omega^{2}_{1}\left[\kappa-\Psi^{2}(t_{I})\right]
\left[1-\Psi^{2}(t_{I})\right]=0
\end{eqnarray} 
with
\begin{eqnarray} 
\dot{\Psi}(t)=\frac{\sqrt{\kappa}\omega_{1}
\displaystyle \sum_{i=1}^{2} \nu_{i}\sinh(\omega_{3-i}t+\theta_{3-i})\cosh(\omega_{i}t+\theta_{i})}
{\sinh^{2}(\omega_{2}t+\theta_{2})}
\end{eqnarray} 
where $\nu_{1}=1$ and $\nu_{2}=-\sqrt{\kappa}$.

In addition, the Hubble function predicted by the UDM model 
can be viewed as the sum of basic hyperbolic functions:
\begin{eqnarray} 
H(t)\equiv \frac{\dot{a}}{a}=\frac{2(x_{2}\dot{x_{2}}-x_{1}\dot{x_{1}})}
{3(x^{2}_{2}-x^{2}_{1})}=\frac{2}{3}\omega_{2}f(t)=H_{0}\frac{f(t)}{f(t_{0})} 
\end{eqnarray} 
with 
\begin{eqnarray} 
f(t)=\frac{\displaystyle \sum_{i=1}^{2}\nu_{i}\sinh{(\omega_{3-i}t+\theta_{3-i})}
\cosh{(\omega_{3-i}t+\theta_{3-i})} }
{\sinh^{2}{(\omega_{2}t+\theta_{2})}-\kappa\sinh^{2}{(\omega_{1}t+\theta_{1})}}
\end{eqnarray} 
where $H_{0}$ is the Hubble constant.
In this work we use $H_0=100h$Kms$^{-1}$Mpc$^{-1}$ 
with $h=0.72$ \cite{Freedman01} or $H_{0}=h/9.778\simeq 0.0736$Gyr$^{-1}$ 
corresponding to $t_{0}\approx H^{-1}_{0}\simeq 13.6$Gyr.
Also we can relate the frequency $\omega_{2}$
of the hyperbolic oscillator in the $x_{2}$ axis with the 
well known cosmological parameters. Indeed, $\omega_{2}$
is given by  
\begin{eqnarray} 
\label{omm} 
\omega_{2}=\frac{3H_{0}\sqrt{1-\Omega_{m}}}{2}  \;\;\;,
\end{eqnarray} 
while $\theta_{1} \propto H_{0}t_{0}$ has no units.
Notice that, $\Omega_{m}$ is the matter density at the present time.

\section{Cosmological constraints and predictions}
In this work we use the so called Baryonic Acoustic Oscillations 
(BAOs) in order to constrain the current cosmological models.
BAOs are produced by pressure (acoustic) waves in the photon-baryon plasma in the 
early universe, generated by dark matter overdensities. First 
evidence of this excess was recently found in the clustering 
properties of the luminous SDSS red-galaxies 
\cite{Eis05}, \cite{Pad07} and it can provide a ''standard ruler'' with which 
we can put constraints 
on the cosmological models.
For a spatially flat FLRW we use the following estimator:
\begin{equation}
A({\bf p})=\frac{\sqrt{\Omega_{m}}}{[z^{2}_{s}H(a_{s})/H_{0}]^{1/3}}
\left[\int_{a_{s}}^{1} \frac{{\rm d}y}{y^{2}H(y)/H_{0}}
\right]^{2/3}
\end{equation}
measured from the SDSS data to be $A=0.469\pm 0.017$, where $z_{s}=0.35$ 
[or $a_{s}=(1+z_{s})^{-1}\simeq 0.75$]. Therefore, 
the corresponding $\chi^{2}_{\rm BAO}$ function is simply written 
\begin{equation}
\chi^{2}_{\rm BAO}({\bf p})=\frac{[A({\bf p})-0.469]^{2}}{0.017^{2}}
\end{equation}
where ${\bf p}$ is a vector containing the cosmological
parameters that we want to fit.

Also, we additionally utilize the sample of 
192 supernovae of Davies et al. \cite{essence}.   
In this case, the $\chi^{2}_{\rm SNIa}$ function becomes:
\begin{equation}
\label{chi22} 
\chi^{2}_{\rm SNIa}({\bf p})=\sum_{i=1}^{192} \left[ \frac{ {\cal \mu}^{\rm th}
(a_{i},{\bf p})-{\cal \mu}^{\rm obs}(a_{i}) }
{\sigma_{i}} \right]^{2} \;\;.
\end{equation}
where $a_{i}=(1+z_{i})^{-1}$ is the observed scale factor of
the Universe, $z_{i}$ is the observed redshift, ${\cal \mu}$ is the 
distance modulus ${\cal \mu}=m-M=5{\rm log}d_{\rm L}+25$
and $d_{\rm L}(a,{\bf p})$ is the luminosity distance 
\begin{equation}
d_{\rm L}(a,{\bf p})=\frac{c}{a} \int_{a}^{1} \frac{{\rm d}y}{y^{2}H(y)}
\end{equation}
where $c$ is the speed of light ($\equiv 1$ here).
Finally we can combine the above probes by using a joint likelihood analysis:
$${\cal L}_{tot}({\bf p})=
{\cal L}_{\rm BAO} \times {\cal L}_{\rm SNIa} \;\;\;\;
\chi^{2}_{tot}({\bf p})=\chi^{2}_{\rm BAO}+\chi^{2}_{\rm SNIa} \;\;\;,$$ 
in order to put even further constraints on the parameter space used. 
Note, that we define the likelihood estimator\footnote{Likelihoods
are normalized to their maximum values.} as:
${\cal L}_{j}\propto {\rm exp}[-\chi^{2}_{j}/2]$.

\subsection{The standard $\Lambda$-Cosmology}
Without wanting to appear too pedagogical, we remind the reader
of some basic elements of the concordance $\Lambda$-cosmology.
In this framework, the normalized scale factor 
of the universe is
\begin{eqnarray} 
\label{all} 
a^{\Lambda}(t)=\left(\frac{\Omega_{m}}{1-\Omega_{m}}\right)^{1/3}
\sinh^{2/3}\omega_{2}t 
\end{eqnarray} 
The Hubble function is written as 
\begin{eqnarray} 
H(t)=\frac{2}{3}\omega_{2}f(t)=
\frac{2}{3}\omega_{2}
\coth\omega_{2}t
\end{eqnarray} 
Comparing the $\Lambda$ model with the observational data 
(we sample $\Omega_{m} \in [0.1,1]$ in steps of
0.01) we find 
that the best fit value is $\Omega_{m}=0.26\pm 0.01$ with 
$\chi_{tot}^{2}(\Omega_{m})\simeq 195$ (${\rm dof}=192$) 
in a very good agreement with the 
5 years WMAP data \cite{Komatsu08}.
The inflection point takes place at 
\begin{eqnarray} 
t^{\Lambda}_{I}=\frac{1}{\omega_{2}}
\sinh^{-1}\left(\frac{1}{2}\right)
  \;\;\;\; a^{\Lambda}_{I}=\left[\frac{\Omega_{m}}{2(1-\Omega_{m})}\right]^{1/3}
\end{eqnarray}  
Therefore, we estimate $\omega_{2}\simeq 1.29H_{0} \simeq 0.095$Gyr$^{-1}$,
$t^{\Lambda}_{I}\simeq 0.51t_{0}$ and $a^{\Lambda}_{I}\simeq 0.56$.
The deceleration parameter at the present time is 
$q_{0}\equiv -\ddot{a}/\dot{a}^{2}|_{a=1}\simeq -0.61$.

 \begin{figure}
         \centerline{\includegraphics[width=20pc] {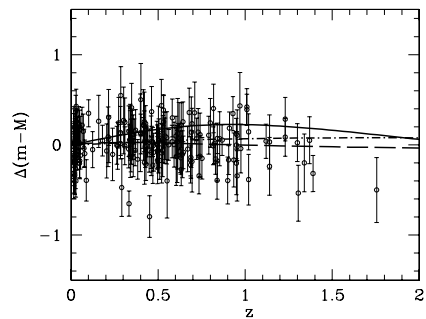}} 
  \caption{
Residual magnitudes (relative to the traditional 
$\Omega_{\Lambda}=0.74$, $\Omega_{m}=0.26$ model) 
of 192 SNIa data (open points)
from \cite{essence} as a function of redshift. For 
comparison we plot the $\Delta(m-M)_{UDM-\Lambda}$
(solid line), $\Delta(m-M)_{CPL-\Lambda}$ (long dashed line) and  
$\Delta(m-M)_{VCG-\Lambda}$ (dot dashed line).
  }
         \label{Fighubmod}
 \end{figure}

\subsection{The parametric Dark Energy model}
In this case we use a simple parameterization for the dark energy
equation of state parameter which is based on a Taylor expansion around the
present time (see Chevallier \& Polarski \cite{chev01} and 
Linder \cite{Lin03}, hereafter CPL) 
\begin{equation}
{\rm w}(a)={\rm w}_{0}+{\rm w}_{1}(1-a) \;\;\;.
\end{equation}
The Hubble parameter is given by:
\begin{equation}
H(a)=H_{0}\left[\Omega_{m}a^{-3}+(1-\Omega_{m})
a^{-3(1+{\rm w}_{0}+{\rm w}_{1})}e^{3{\rm w}_{1}(a-1)} \right]^{1/2} 
\end{equation}
where ${\rm w}_{0}$ and ${\rm w}_{1}$ are constants. 
We sample the unknown parameters as follows: ${\rm w}_{0} \in [-2,-0.4]$ 
and ${\rm w}_{1} \in [-2.6,2.6]$ in steps of 0.01. 
We find that for $\Omega_{m}=0.26$ the overall likelihood function 
peaks at ${\rm w}_{0}=-1.20^{+0.28}_{-0.20}$ 
and ${\rm w}_{1}=1.14^{+1.0}_{-1.9}$ 
while the corresponding $\chi_{tot}^{2}({\rm w}_{0},{\rm w}_{1})$ is 193.6
(${\rm dof}=191$). 
The deceleration parameter at the present time is 
$q_{0}\simeq -0.83$.

\subsection{The Variable Chaplygin Gas as an alternative to dark energy}
Let us consider now a completely different model namely the
variable Chaplygin gas (herafter VCG) which corresponds 
to a Born-Infeld tachyon action \cite{Bento03,Guo05}. Recently, an 
interesting family of Chaplygin gas models was found to be consistent
with the current observational data \cite{Vcgdata}. 
In the framework of a 
spatially flat FLRW metric, 
it can be shown that the Hubble function takes the following formula:
\begin{equation}
H(a)=H_{0}\left[\Omega_{b}a^{-3}+(1-\Omega_{b})
\sqrt{B_{s}a^{-6}+(1-B_{s})a^{-n}} \right]^{1/2}
\end{equation}  
where $\Omega_{b}\simeq 0.021h^{-2}$ is the density 
parameter for the baryonic matter \cite{Kirk03} and 
$B_{s} \in [0.01,0.51]$ in steps of
0.01 and $n\in [-4,4]$ in steps of 0.02.
The corresponding 
effective equation of state parameter ${\rm w}(a)=P_{DE}/\rho_{DE}$ 
is related to $H(a)$ according to
\begin{equation}
{\rm w}(a)=\frac{-1-\frac{2}{3}a\frac{dlnH}{da}}
{1-(\frac{H_{0}}{H})^{2}\Omega_{m}a^{-3}} \;\;\;,
\end{equation}  
while the effective matter density parameter is: 
$\Omega^{eff}_{m}=\Omega_{b}+(1-\Omega_{b})\sqrt{B_{s}}$.
We find that the best fit parameters are 
$B_{s}=0.07\pm 0.02$ and $n=1.06\pm 0.33$ ($\Omega^{eff}_{m}\simeq 0.29$)
with $\chi_{tot}^{2}(B_{s},n)=193.7$ (${\rm dof}=191$) and 
the present value of the deceleration parameter is 
$q_{0} \simeq -0.60$.

\subsection{The UDM comparison with other Dark energy models}
In order to predict analytically the time evolution of the main cosmological
functions [$\phi(t)$, $a(t)$, $H(t)$ and ${\rm w}(t)$] we have to define 
the corresponding unknown constants of the problem
($\omega_{1},\theta_{1},\omega_{2},\theta_{2}$).
At the same time, from the restrictions found in section 3 (see
eqs. \ref{om11}, \ref{con2} and \ref{omm}), we can reduce the parameter space to 
$(\Omega_{m},\theta_{1})$. We do so by fitting the predictions
of the UDM cosmological model and recent observational data.
Here, we use $\Omega_{m} \in [0.1,1]$ and $\theta_{1}\in [-1,0]$ in steps
of 0.01.

Figure 1 (thin dashed lines) shows
the 1$\sigma$, 2$\sigma$ and 3$\sigma$
confidence levels in the 
$(\Omega_{m},|\theta_{1}|)$ plane when using BAOs. Obviously, 
the $\theta_{1}$ parameter 
is not constrained by this analysis
and all the values in the interval $-1\le \theta_{1}\le 0$
are acceptable. However, the BAOs statistical analysis puts constraints 
on the matter density parameter $\Omega_{m} \simeq 0.25$. 

Therefore, in order to put further constraints on 
$\theta_{1}$ we additionally utilize the SNIa data.
In figure 1 (thick solid lines), we present the 
SNIa likelihood contours
and we find that the best fit solution is $\Omega_{m}\simeq 0.4$
and $\theta_{1}\simeq -0.05$.
The joint likelihood function peaks at 
$\Omega_{m}=0.25^{+0.02}_{-0.01}$ and $\theta_{1}=-0.39^{+0.04}_{-0.08}$
($\theta_{2}\simeq 0.54$) with
$\chi_{tot}^{2}(\Omega_{m},|\theta_{1}|)
\simeq 194.1$ (${\rm dof}=191$). 
Note that the errors of the fitted parameters represent $1\sigma$ 
uncertainties.
In the insert plot of figure 1 we provide the solutions 
(circles-BAOs and triangles-SNIa) within 
$1\sigma$ contours in the $(\Omega_{m},\phi_{0})$ 
plane, where $\phi_{0}$ is the present value of the scalar field.
The corresponding best fit value of the scalar field 
(see eqs. \ref{psi11} and \ref{phi11})
is $\phi_{0}\simeq 0.42$ or $0.084$ in Planck units 
($G=c\equiv 1$), while the frequencies are 
$\omega_{1}\simeq 0.067$Gyr$^{-1}$ and 
$\omega_{2}\simeq 0.095$Gyr$^{-1}$.
It is interesting to mention that although  
the frequency ($\omega_{1}\sim 0.9H_{0}$) 
of the hyperbolic oscillator in the $x_{1}$ axis is somewhat 
less than the present 
expansion rate of the universe,
the $\omega_{2}$ is equal to the value predicted by 
the $\Lambda$ cosmology (see section 4.1).

 \begin{figure}
         \centerline{\includegraphics[width=20pc] {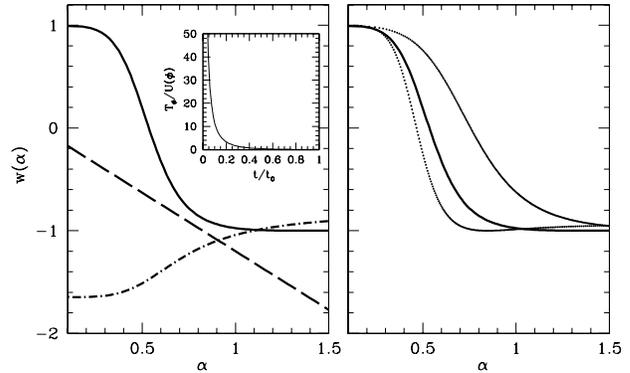}} 
  \caption{{\it Left Panel:}The equation of state parameter as a function
of the scale factor of the Universe. 
The lines corresponds to UDM (solid), CPL (long dashed) 
and VCG (dot dashed).
In the insert panel we present the 
time evolution of $T_{\phi}/V(\phi)$. Note, that $T_{\phi}$ is the kinetic 
energy of the scalar and $V(\phi)$ is the potential.
{\it Right Panel:} The functional form of the equation of state
parameter for various UDM models. The upper line corresponds
to $(\Omega_{m},\theta_{1})=(0.73,-1)$ while the bottom line 
corresponds to $(\Omega_{m},\theta_{1})=(0.19,-0.1)$.
Note that the solid thick line corresponds to the best-fit
parameters $(\Omega_{m},\theta_{1})=(0.25,-0.39)$. We find that
initially all the UDM models [$\Omega_{m} \in (0.19,0.73)$, 
$\theta_{1} \in (-1,-0.1)$]
start from ${\rm w}\longrightarrow +1$ and 
they reach ${\rm w} \simeq O(-1)$ close 
to the present time.} 

         \label{Figw11}
 \end{figure}

Knowing now the parameter space  
($\omega_{1},\theta_{1},\omega_{2},\theta_{2}$)
we investigate, in more detail, the 
correspondence of the UDM model with the different dark energy models 
(see sections 4.1, 4.2 and 4.3) in order to show
the extent to which they compare. 
Our analysis provides an evolution of the UDM  
scale factor seen in the upper panel 
of figure 2 as the solid line, which closely resembles, especially
at late times ($0.6<a\le 1.5$), the
corresponding scale factor of the 
$\Lambda$ (short dashed), VCG (dot dashed),
and CPL (long dashed). Note that the  
UDM deceleration parameter at the present time is $q_{0}\simeq -0.62$.
However, for $a>1.5$, the
CPL and the VCG scale factors 
evolve more rapidly than the other two models (UDM and
$\Lambda$ cosmology). Also
it is clear that
an inflection point [$\ddot{a}(t_{I})=0$] is present in 
the evolution of the UDM 
scale factor. The UDM inflection point is located at 
$t_{I}\simeq 0.46t_{0}$ which corresponds to $a_{I}\simeq 0.61$
and is somewhat different than the value predicted from the 
usual $\Lambda$ cosmology (see section 4.1). Before the inflection point,
the UDM appears to be more decelerated from the other three dark energy cosmological models
due to the fact that the second term [$\propto {\rm sinh}^{2}(\omega_{1}t+\theta_{1}$)]
in eq.(\ref{a11}) plays an important role.
From figure 2 it becomes clear that the UDM model 
reaches a maximum deviation from the other three dark energy models
prior to $a \sim 0.15$ ($z \sim 5.5$).
In order to investigate whether the expansion of the observed universe
follows such a possibility we need a visible
distance indicator (better observations) at redshifts $z>2$.
 
The evolution of the scalar field is presented in the
bottom panel of figure 2, while in the insert figure we plot 
the scalar field dependence of the potential energy normalized 
to unity at the present time. 
As we have stated in section 3.1 there is one minimum at $\phi=0$ that
corresponds to $t_{m}=-\theta_{1}/\omega_{1}\sim 0.4t_{0}$.
To conclude, we plot in figure 3 the relative deviations of the distance modulus, 
$\Delta(m-M)$, of the dark energy models used here 
from the traditional $\Lambda$ cosmology. 
Notice that the open points represent the following deviation: $(m-M)_{SNIa}-(m-M)_{\Lambda}$. 
Within the SNIa redshift range $0.016\le z \le 1.775$ ($0.360\le a \le 0.984$) 
the VCG distance modulus is close to the $\Lambda$ one.
The largest deviations of the distance moduli occur at redshifts around 
0.5-1 for the UDM and 1.1-1.5 for the CPL model respectively.
 
\subsection{The equation of state parameter}
We would like to end this section with a 
discussion on the dark energy equation of state.
As we have stated already in the introduction, there is a possibility 
for the equation of state parameter to be a function of time 
rather than a constant ratio between the pressure
and the energy density. Within the framework of the scalar field cosmology 
the equation of state parameter is derived from the field model  
and in general it is a complicated function of time, even when the
potential is written as a simple function of the scalar field.
In our case we have  
\begin{eqnarray} 
{\rm w}(t)=\frac{P_{\phi}}{\rho_{\phi}}=\frac{{\dot \phi}^{2}-2V(\phi)}{{\dot \phi}^{2}+2V(\phi)}
\end{eqnarray}
or else (see eqs.\ref{phi11}, \ref{potent11}) 
\begin{eqnarray} 
{\rm w}(t)=\frac{{\dot \Psi}^{2}(t)-\omega^{2}_{1}[\kappa-\Psi^{2}(t)][1-\Psi^{2}(t)]}
{{\dot \Psi}^{2}(t)+\omega^{2}_{1}[\kappa-\Psi^{2}(t)][1-\Psi^{2}(t)]} \;\;.
\end{eqnarray}
Note that $\Lambda-$models can be described by scalar models with
${\rm w}$ strictly equal to -1. 
Using our best fit parameters we present in left panel of figure 4, 
the equation of state parameter as a function of the scale factor
for the different dark energy models. The UDM model (solid line) is the 
only case that provides positive values for the equation of 
state parameter at early epochs.  
We have checked the UDM scenario against the cosmic coincidence problem 
(why the matter energy density and the dark energy density are of the same
order at the present epoch) by utilizing the basic tests proposed by
\cite{stein99}. These are: (a) at early enough 
times the equation of state parameter tends to its maximum value, 
${\rm w}\longrightarrow +1$, which means that the dark energy density 
initially takes large values. So 
as long as the scalar field rolls down the potential 
energy, $V(\phi)$, decreases rapidly and the kinetic energy 
$T_{\phi}={\dot \phi}^{2}/2$ takes a large value, 
(b) then $\phi$ continues to roll down, the dark energy density
decreases and the equation of state parameter remains close to unity for 
a quite long period of time ($a<0.2$) and (c) 
for $0.2\le a\le 0.95$ the equation of state parameter 
is a decreasing function of time and it 
becomes negative at $a>0.56$.
Before that epoch, the potential 
energy of the scalar field remains less than the kinetic energy 
(see the insert plot in the left panel of figure 4) and
the equation of state parameter 
(or the scalar field) resembles background matter. 
In a special case where ${\rm w}=0$ [or $T_{\phi}\simeq V(\phi)$]
the equation of state behaves exactly like that of pressure-less matter.
For ${\rm w}=-1/3$ we reach the same expansion as in an open universe, 
because the dark energy density evolves as $a^{-2}$ 
and has no effect on $\ddot{a}$. 
In fact, we verify that prior to the inflection point 
${\rm w}(t_{I})\simeq -0.334$, which means that after $t_{I}$ 
the accelerating expansion of the Universe starts. 
Finally, ${\rm w}\simeq -1$ close to the present epoch 
$a \sim 1$ and the scalar field is effectively
frozen (the same situation seems to hold also in the 
limit $a \gg 1$). 
This is to be expected because at this period the scalar field varies slowly
with time (see the insert panel of figure 4), so that $T_{\phi} \ll V(\phi)$ 
and the dark energy fluid asymptotically reaching the de-Sitter regime
(cosmological constant).
 \begin{figure}
         \centerline{\includegraphics[width=20pc] {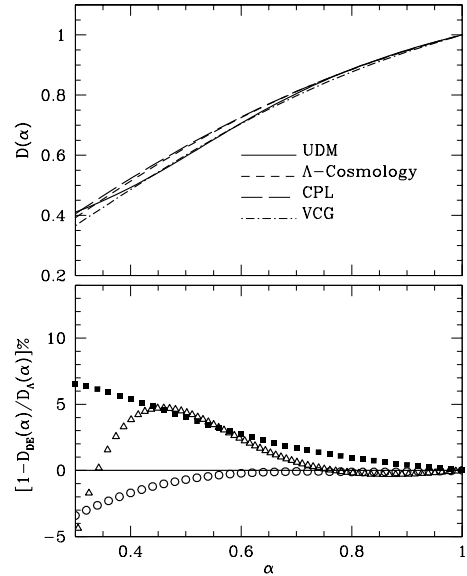}} 
  \caption{{\it Upper Panel:} The evolution of the growth factor for different 
dark energy models. 
The lines correspond to UDM (solid), VCG (dot dashed),
$\Lambda$ (short dashed) and CPL (long dashed) models.
{\it Bottom Panel:} The deviation 
$(1-D_{DE}/D_{\Lambda})\%$ of the growth factor for various 
dark energy models with respect to the $\Lambda$ solution.
The points represent the comparison: (a) UDM-$\Lambda$ (open triangles), 
(b) VCG-$\Lambda$ (solid squares) and (c) CPL-$\Lambda$ (open circles).}
         \label{Figcondz}
 \end{figure}

In order to conclude this discussion, it is interesting 
to point out that we also investigate
the sensitivity of the above results to the matter density parameter. 
As an example, in the right panel of figure 4 we present 
the evolution of the equation of state parameter 
for $(\Omega_{m},\theta_{1})=(0.73,-1)$
[upper line] and $(\Omega_{m},\theta_{1})=(0.19,-0.1)$ [bottom line].
We confirm that in the range 
$\Omega_{m} \in (0.19,0.73)$ and $\theta_{1} \in (-1,-0.1)$
the general behavior (described before)
of the functional form of the equation 
of state parameter is an intermediate case between the above lines
for $a\le 1$ and thus it depends weakly on the
values of the parameter space ($\Omega_{m},\theta_{1}$). 
Therefore, our main cosmological results for the UDM scenario persist for all
physical values of $\Omega_{m}$ and it strongly indicates that the UDM
model over-passes the cosmic coincidence problem.

\section{Evolution of matter perturbations}
In this section we attempt to study the dynamics at 
small scales by generalizing the basic linear and non-linear equations
which govern the behavior of the matter perturbations
within the framework of a UDM flat cosmology.
Also we compare our predictions with those found for the dark energy 
models used in this work (see sections 4.1, 4.2 and 4.3).
This can help us  
to understand better the theoretical expectations of
the UDM model as well as the variants from the other dark energy models.

\subsection{The Evolution of the linear growth factor}
The evolution equation 
of the growth factor for models where the dark energy
fluid has a vanishing anisotropic stress and the matter fluid is not
coupled to other matter species is given by
(\cite{Peeb93}, \cite{Stab06}, \cite{Uzan07}): 
\begin{equation}
\frac{{\rm d}^{2}D}{{\rm d}N^{2}}+\left(2+\frac{1}{H}\frac{{\rm d}H}
{{\rm d}N}\right)
\frac{{\rm d}D}{{\rm d}N}-\frac{3}{2}\Omega_{m}(a)D=0
\label{deltatime1} 
\end{equation}
where $N={\rm ln}a$ and $\Omega_{m}(a)=\Omega_{m}a^{-3}H^{2}_{0}/H^{2}(a)$.
Useful expressions of the growth factor can be found for the
$\Lambda$CDM cosmology in \cite{Peeb93}, for 
the quintessence scenario (${\rm w}=const$) in 
\cite{Silv94}, \cite{Wang98}, \cite{Bas03}, \cite{Nes08}, 
for dark energy models with 
a time varying equation of state in 
\cite{Linca08} and for the 
scalar tensor models in \cite{Gann08}.
In the upper panel of figure 5 we present the growth factor 
evolution which is derived by solving
numerically eq. (\ref{deltatime1}), for the four dark energy models 
(including the UDM). Note that the growth factors are normalized to unity 
at the present time. The behavior of the UDM growth factor
(solid line) has the expected form, i.e. it is an increasing function 
of the scale factor. Also we find that the growth factor in the UDM model
is almost an intermediate case between the VCG (dot dashed line) and 
CPL (long dashed line) models respectively. 
In the bottom panel of figure 5 we
show the deviation, $(1-D_{DE}/D_{\Lambda})\%$, of the growth factors
$D_{DE}(a)$ for the current dark energy models
with respect to the $\Lambda$ solution $D_{\Lambda}(a)$. 
Assuming now that clusters have formed prior to
the epoch of $z_{\rm f}\simeq 1.4$ ($a_{\rm f}\sim 0.42$), in which 
the most distant cluster has been found 
\cite{Mul05},
the UDM scenario (open triangles) deviates from the $\Lambda$ solution by $4.2\%$ while
the CPL (open circles) and VCG (solid squares) 
deviates by $-1.5\%$ and $5.1\%$ respectively.
Also at the $\Lambda$-inflection point ($a^{\Lambda}_{I}\simeq 0.56$)
we find the following results: (i) UDM-$\Lambda$ $3.3\%$,
(ii) CPL-$\Lambda$ $-0.4\%$ and (iii) VCG-$\Lambda$ $3.2\%$.
To conclude this discussion it is obvious that 
for $a \ge 0.7$ the UDM growth factor tends to the $\Lambda$ solution
(the same situation holds for the CPL model but with $a \ge 0.55$).

 \begin{figure}
         \centerline{\includegraphics[width=20pc] {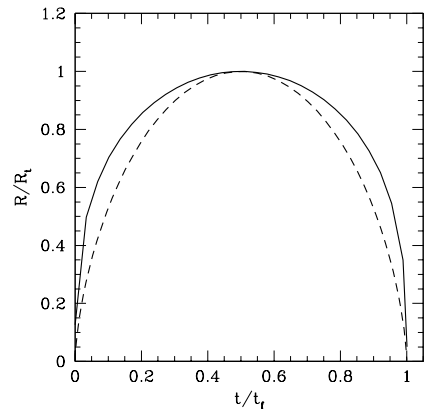}} 
  \caption{Evolution of radius of a collapsing overdense region.
The solid and the dashed line corresponds to the UDM ($\Omega_{m}=0.25$)
and $\Lambda$ cosmology ($\Omega_{m}=0.26$) respectively.}
         \label{Figcondz1}
 \end{figure}

\subsection{The spherical collapse model}
The so called spherical collapse model, which has a long history
in cosmology, is a simple but still a fundamental tool 
for understanding how a small spherical patch [with radius $R(t)$] of
homogeneous overdensity forms a bound system via 
gravitation instability \cite{Gunn72}. 
From now on, we will call $a_{\rm t}$ the scale factor of the
universe where the overdensity reaches its maximum expansion ($\dot R=0$)
and $a_{\rm f}$ the scale factor in which the sphere virializes,
while $R_{\rm t}$ and $R_{\rm f}$ the corresponding radii of the spherical
overdensity. Note that in the spherical region, 
$\rho_{mc}\propto R^{-3}$ is the matter density while 
$\rho_{\phi_{c}}$ will denote the corresponding density of the dark energy. 
In order to address the issue of the
dark energy in the gravitationally bound systems (clusters of galaxies) 
we can consider the following assumptions: (i)
clustered dark energy considering that that the whole system virializes
(matter and dark energy), (ii) the dark energy remains clustered
but now only the matter virializes
and (iii) the dark energy remains homogeneous and only 
the matter virializes (for more details 
see \cite{Mota04}, \cite{Maor05} \cite{Wang06} and \cite{Basi07}).  
Note, that in this work we are using the third possibility.

Here we review only some basic concepts of the problem
based on the assumption that the dark energy component under a 
scale of galaxy clusters can be treated as being homogeneous:
$\rho_{\phi_{c}}(t)=\rho_{\phi}(t)$, $\phi_{c}(t)=\phi(t)$ 
and ${\rm w}_{c}(t)={\rm w}(t)$.
In general the evolution of the spherical perturbations as the latter decouple 
from the background expansion is given by the Raychaudhuri equation:
\begin{equation}
3\ddot{R}=-4\pi GR[\rho_{mc}+\rho_{\phi_{c}}(1+{\rm w}_{c})]\;\; 
{\rm here}\;\;4\pi G\equiv 1/2.
\end{equation}  
Now within the cluster region the evolution of the dark energy 
component is written as (see \cite{Mota04})
\begin{equation}
\dot{\rho}_{\phi_{c}}+3\frac{\dot R}{R}(1+{\rm w}_{\phi_{c}})\rho_{\phi_{c}}
=\Gamma 
\end{equation} 
while if we consider a scalar field the above equation becomes
\begin{equation}
      \ddot{\phi}_{c}+3 \frac{\dot{R}}{R}\dot{\phi}_{c}+U^{\prime}(\phi_{c})=
\frac{\Gamma}{\dot \phi}
\end{equation} 
where
\begin{equation}
\Gamma=-3\left(\frac{\dot{a}}{a}-\frac{\dot{R}}{R} \right) 
\dot{\phi}^{2}_{c}\;\;\;.
\end{equation} 
Figure 6 presents examples of $R(t)$ obtained for the UDM (solid line)
and for the concordance $\Lambda$ model (dashed line). The time needed 
for a spherical shell to re-collapse 
is twice the turn-around time, $t_{\rm f}\simeq 2t_{\rm t}$. 

On the other hand, utilizing both the virial theorem and the energy 
conservation we reach to the following condition:
\begin{equation}
\left[\frac{1}{2}R\frac{\partial}{\partial R} (U_{G}+U_{\phi_{c}})+
U_{G}+U_{\phi_{c}}\right]^{a=a_{\rm f}}=
\left[U_{G}+U_{\phi_{c}} \right]^{a=a_{\rm t}} 
\label{virial} 
\end{equation} 
where $U_{G}=-3GM^{2}/5R$ is the potential energy and 
$U_{\phi_{c}}=-4\pi GM(1+3{\rm w}_{c})\rho_{\phi_{c}}R^{2}/5$
is the potential energy associated with the dark energy for the spherical
overdensity (see \cite{Mota04} and \cite{Maor05}; in our 
case $4\pi G\equiv 1/2$). 
Using the above formulation we can obtain a cubic equation
that relates the ratio between the virial $R_{\rm f}$ 
and the turn-around outer radius $R_{\rm t}$
the so called 
collapse factor ($\lambda=R_{\rm f}/R_{\rm t}$).   
Notice that 
eq.(\ref{virial}) is valid when the ratio of the system's dark energy 
to the matter's densities at the time of the 
turn-around takes relatively small values \cite{Wang06}. 
Of course in the case of ${\rm w}_{c}=-1$ the above expressions
get the usual form for $\Lambda$ cosmology (\cite{Bas03}, \cite{lahav91})
 while for an Einstein-de Sitter model ($\Omega_{m}=1$)
we have $\lambda=1/2$.   
 \begin{figure}
         \centerline{\includegraphics[width=20pc] {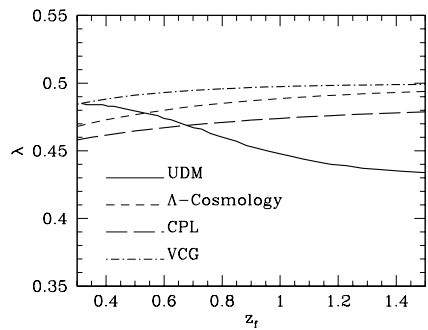}} 
  \caption{The collapse factor versus the redshift of virialization
for various dark energy models.}
         \label{Figcondz1}
 \end{figure}
Finally solving numerically eq.(\ref{virial}) [it can be done also analytically] 
we calculate the collapse factor. In particular, 
figure 7 shows the behavior of the collapse factor for the current 
cosmological models starting from the UDM (solid line), 
$\Lambda$ (short dashed), VCG (dot dashed),
and CPL (long dashed). We find that the collapse factor
lies in the range $0.43\le \lambda \le 0.50$ in agreement with 
previous studies (
\cite{Bas03}, \cite{Mota04}, \cite{Maor05}, \cite{Wang06}, \cite{Horel05},
\cite{Perc05}, \cite{Basi07}). 
Prior to the cluster formation epoch
($z_{\rm f}\simeq 1.4$) 
the UDM scenario, appears to produce more bound systems with respect to 
the other dark energy models. Indeed, we find the following values: 
$\lambda_{UDM}\simeq 0.44$, $\lambda_{\Lambda}\simeq 0.49$, 
$\lambda_{CPL}\simeq 0.48$ and $\lambda_{VCG}\simeq 0.50$.  
Also it becomes clear that the UDM collapse factor decreases slowly with 
the redshift of virialization $z_{f}$, 
due to its positive equation of state parameter. This is also
incorporated by the fact that at early epochs 
the cosmic expansion of the UDM model 
is much more decelerated than in the other three
dark energy models. The latter 
result is in agreement with those obtained by  
\cite{Mota04}. They 
found a similar behavior for the collapse factor
by considering several potentials with an exponential phase.

\section{Conclusions}
In this work we investigate analytically and numerically 
the large and small scale dynamics of the scalar field FLRW flat cosmologies in the 
framework of the so called {\it Unified Dark Matter} scenario. In particular using a Hamiltonian 
formulation we find that the 
time evolution of the basic cosmological functions are described in terms of hyperbolic functions. 
This theoretical approach yields analytical solutions which  
can accommodate a late time accelerated
expansion, equivalent to either the dark energy or the standard $\Lambda$ models.
Furthermore, based on a joint likelihood analysis using
the SNIa data and the Baryonic Acoustic Oscillations, we put
tight constraints on the main cosmological parameters 
of the UDM cosmological model. In particular we find 
$\Omega_{m}\simeq 0.25$ and the scalar field at the present time
is $\phi_{0}\simeq 0.42$ or 0.084 (in Planck units). 
Also, we compare the UDM scenario with various 
dark energy models namely $\Lambda$ cosmology,
parametric dark energy model and variable Chaplygin gas. 
We find that the cosmological behavior of the UDM scalar field model 
is in a good agreement, especially after the inflection point,
with those predicted by the above dark energy models
although there are some differences especially at early epochs.
In particular, we reveal
that the UDM scalar field cosmology has three important differences 
over the other three dark energy models considered:
\begin{itemize}

\item It can pick up positive values of the equation of state 
parameter at large redshifts ($z>0.8$). Also, 
it behaves relatively well with respect to 
the cosmic coincidence problem.

\item At early enough epochs ($a \sim 0.15$ or $z\sim 5.5$) 
the cosmic expansion in the UDM model 
is much more decelerated than in the other three
dark energy models. 
In order to investigate whether the expansion of the observed universe
has the above property, we need a visible
distance indicator (better observations) at 
high redshifts ($2\le z \le 6$).

\item Close to the cluster formation epoch, 
its collapse factor $\lambda_{UDM}$ is less
than $12\%$ of the corresponding factor of the other three
dark energy models. This feature points to the direction 
that perhaps the $\lambda$ parameter can be used as a 
cosmological tool.

\end{itemize}

\begin{acknowledgments}
We thank prof. George Contopoulos, Dr. Manolis Plionis 
and the anonymous referee for their 
useful comments and suggestions. 
G. Lukes-Gerakopoulos was supported by the Greek Foundation 
of State Scholarships (IKY).
\end{acknowledgments}


\begin{thebibliography} {plain}

 \bibitem {Riess07}
 A. G. Riess, {\em et al.}, Astrophys. J., {\bf 659}, 98, (2007)

\bibitem{essence} W.M. Wood-Vasey {\em et al.}, 
 Astrophys. J., {\bf 666}, 694, (2007);
T.M. Davis {\em et al.}, Astrophys. J., {\bf 666}, 716, (2007)

 \bibitem {Spergel07}
 D. N. Spergel, {\em et al.}, Astrophys. J. Suppl., {\bf 170}, 377, (2007)

\bibitem{Komatsu08}
E. Komatsu, {\em et al.}, Astrophys. J. Suppl., submitted, (2008), [arXiv:0803.0547]

 \bibitem {Ratra88}
B. Ratra, P. J. Peebles, Phys. Rev D., {\bf 37}, 3406, (1988)
 \bibitem {Weinberg89}
 S. Weinberg, Rev. Mod. Phys., {\bf 61}, 1, (1989)

\bibitem{Wetterich:1994bg}
C.~Wetterich, Astron. Astrophys. {\bf 301}, 321 (1995)

 \bibitem{Caldwell98}
R. R. Caldwell, R. Dave, P. J. Steinhardt, Phys. Rev. Lett., {\bf 80}, 1582, (1998) 

\bibitem{KAM}
A. Kamenshchik, U. Moschella, V. Pasquier, Phys. Lett. B., {\bf 511}, 265, (2001)

\bibitem{Caldwell}
R.R., Caldwell,  E. V., Linder, Phys. Rev. Lett., {\bf 95}, 141301, (2005)

 \bibitem {Peebles03}
P. J. Peebles, B. Ratra, Rev. Mod. Phys., {\bf 75}, 559, (2003)

\bibitem{Brax:1999gp}
P.~Brax, J.~Martin, Phys. Lett. {\bf B468}, 40 (1999)

\bibitem{fein02}
A. Feinstein, Phys. Rev. D., {\bf 66}, 063511, (2002)

\bibitem{chime04}
L. P. Chimento, A. Feinstein, Mod. Phys. Lett. A, {\bf 19}, 761, (2004)

\bibitem{Brookfield:2005td}
A.~W. Brookfield, C.~van~de Bruck, D.~F. Mota, D.~Tocchini-Valentini,
Phys. Rev. Lett. {\bf 96}, 061301 (2006)

\bibitem{Copel06}
 E. J. Copeland, M. Sami, S. Tsujikawa, Int. J. Mod. Phys. D, 
{\bf 15}, 1753 (2006)

\bibitem{Boehmer:2007qa}
C.~G. Boehmer, T.~Harko, Eur. Phys. J. {\bf C50}, 423 (2007)

\bibitem{Friem08}
J. Frieman, M. Turner, D. Huterer, Annual Rev. of Astron. and Astrop.,
 submitted, [arXiv:0803.0982]

 \bibitem{Ozer87}
 M. Ozer, O. Taha, Nucl. Phys. B., {\bf 287}, 776, (1987) 

 \bibitem {Peebles88}
 P. J. Peebles, B. Ratra, Astrophys. J., {\bf 325}, L17, (1988)

 \bibitem {Turner97}
 M. S. Turner, M. White, Phys. Rev. D, {\bf 56}, R4439, (1997)

 \bibitem{Padm03}
 T. Padmanabhan, Phys. Rep., {\bf 6}, 235, (2003)

 \bibitem {Dolgov90}
A. D. Dolgov, M. V. Sazhin, Y. B. Zeldovich, 'Basics of Moderm Cosmology', 
Editions Frontieres, (1990)

\bibitem{sahni00}
V. Sahni, L. Wang, Phys. Rev. D., {\bf 62}, 103517, (2000)

\bibitem{santi00}
D., I. Santiago, A. S. Silbergleit, Phys. Lett. A., {\bf 268}, 69, (2000)

\bibitem{sen02}
A.A. Sen, S. Sethi, Phys. Lett. B., {\bf 532}, 159, (2002)

\bibitem{kehagias04}
A. Kehagias, G. Kofinas, Class. Quantum Grav., {\bf 21}, 3871, (2004)

\bibitem {Gorini05}
V. Gorini, A. Kamenshchik, U. Moschella, V. Pasquier, A. Starobinsky,
Phys. Rev. D, {\bf 72}, 103518, (2005)

\bibitem {Bertacca07}
 D. Bertacca, S. Matarrese, M. Pietroni M., Mod. Phys. Lett. A, {\bf 22}, 2893, (2007)
\bibitem {Kam01}
A. Y. Kamenshchik, U. Moschella, V. Pasquier, Phys. Lett. B., {\bf 511}, 265, (2001)

\bibitem {Bili02}
N. Bilic, G. B. Tupper, R. D. Viollier, Phys. Lett. B., {\bf 535}, 17, (2002)

\bibitem {Taka06}
F. Takahashi, T. T. Yanagida, Phys. Lett. B., {\bf 635}, 57, (2006)
\bibitem{Eis05}
D. J. Eisenstein, et al., Astrophys. J., {\bf 633}, 560, (2005)

\bibitem{Pad07}
N. Padmanabhan, et al., Mon. Not. Roy. Astron. Soc., {\bf 378}, 852, (2007) 

 \bibitem {Page84}
D. N. Page, Class. Quant. Grav., {\bf 1}, 427, (1984)

\bibitem{wetter}
C. Wetterich, Nucl. Phys., B., {\bf 302}, 668, (1988)

\bibitem{zlatev}
I. Zlatev, L. M. Wang, P.J. Steinhardt, Phys. Rev. Lett., {\bf 82}, 896, (1999)

\bibitem {Toporensky06}
 V. A. Toporensky, Symmetry, Integrability and 
Geometry: Methods and Applications, {\bf 2}, 37, (2006)

\bibitem {Turner83}
 M. S. Turner, Phys. Rev. D, {\bf 28}, 1243, (1983)

\bibitem {mata85}
F. Lucchin, S Matarrese, Phys. Rev. D, {\bf 32}, 1316, (1985)


\bibitem{stein99}
P. J. Steinhardt, L. Wang, I. Zlatev, Phys. Rev. D., {\bf 59}, 123504, 
(1999)

\bibitem {Lukes08}
G. Lukes-Gerakopoulos, S. Basilakos, G. Contopoulos,
Phys. Rev. D, {\bf 77}, 043521, (2008)

\bibitem {Gorini04}
V. Gorini, A. Kamenshchik, U. Moschella, V. Pasquier, 
Phys. Rev. D, {\bf 69}, 123512, (2004)

 \bibitem {Scherrer04}
R. J. Scherrer, Phys. Rev. Lett., {\bf 93}, 011301, (2004)

\bibitem {Freedman01}
W. L. Freedman, Astrophys. J., {\bf 553}, 47, (2001)

\bibitem{chev01}
  M., Chevallier, \& D. Polarski,
  Int.\ J.\ Mod.\ Phys.\ D, {\bf 10}, 213, (2001)

\bibitem{Lin03}
  V. E., Linder, Phys. Rev. Lett., {\bf 90}, 091301, (2003)

\bibitem{Bento03}
M. C. Bento, O. Bertolami, A. A. Sen, Phys. Rev. D., {\bf 70}, 083519, (2004)
 
\bibitem{Guo05}
Z. K. Guo, \&, Y. Z., Zhang, astro-ph/0506091, (2005); 
Z. K. Guo, \&, Y. Z., Zhang, astro-ph/050979, (2005)

\bibitem{Vcgdata}
M. Makler, S. Q. de Oliveira, I. Waga, Phys. Lett B., {\bf 555}, 1, (2003);
M. C. Bento, O. Bertolami, A. A. Sen, Phys. Lett. B., {\bf 575}, 172, (2003);
A. Dev, J. S. Alcaniz, D. Jain, Phys. Rev. D., {\bf 67}, 023515, (2003);
Y. Gong, C. K. Duan, Mon. Not. Roy. Astron. Soc., {\bf 352}, 847, (2004); 
Z. H. Zhu, Astron. Astrophys., {\bf 423}, 421, (2004)
L. Amendola, I. Waga, F. Finelli, astro-ph/0509099;  
\bibitem{Kirk03}
D. Kirkman, D. Tytler, N. Suzuki, N., J. M. O'Meara, D. Lubin, 
Astrophys. J. Suppl., ApJS, {\bf 149}, 1, (2003)

\bibitem {Peeb93}
P. J. E. Peebles, 1993, Principles of Physical Cosmology,
Princeton University Press, Princeton New Jersey, (1993)

\bibitem {Stab06}
F. H. Stabenau, \& B. Jain, Phys. Rev. D, {\bf 74}, 084007, (2006)

\bibitem {Uzan07}
P. J.  Uzan, Gen.\ Rel.\ Grav., {\bf 39}, 307, (2007)

\bibitem{Silv94}
V. Silveira, \& I. Waga, Phys. Rev. D., {\bf 64}, 4890, (1994)

\bibitem{Wang98} 
L.  Wang, \& J. P. Steinhardt, Astrophys. J., {\bf 508}, 483, (1998)

\bibitem{Bas03} 
S. Basilakos, Astrophys. J., {\bf 590}, 636, (2003)

\bibitem{Nes08}
S. Nesseris \&  L. Perivolaropoulos, Phys. Rev D., {\bf 77}, 3504, (2008)
 
\bibitem{Linca08} 
V. E. Linder \& N. R. Cahn, Astroparticle Physics, {\bf 28}, 481,
(2007)

\bibitem{Gann08}
R. Gannouji, \&, D. Polarski,  (2008), [arXiv:0802.4196]

\bibitem{Gunn72}
J. E. Gunn, \&, J. R. Gott, Astrophys. J., {\bf 176}, 1, (1972)

\bibitem{Mota04}
D. F. Mota \&, C. van de Bruck C., Astronomy \& Astrophysics, 
{\bf 421}, 71, (2004)

\bibitem{Maor05}
I. Maor, \&, O. Lahav, Journal of Cosmology and
  Astroparticle Physics, {\bf 7}, 3, (2005) 

\bibitem{Wang06}
P. Wang, Astrophys. J., {\bf 640}, 18, (2006)

\bibitem{Basi07}
S. Basilakos, N. Voglis. Mon. Not. Roy. Astron. Soc., {\bf 374}, 269, (2007)

\bibitem{lahav91}
O. Lahav, P. B. Lilje, R. J. Primack, \& M. J. Rees, 
Mon. Not. Roy. Astron. Soc., {\bf 251}, 128, (1991) 

\bibitem{Horel05}
C. Horellou \& J. Berge, Mon. Not. Roy. Astron.  Soc., {\bf 360}, 1393, (2005) 

\bibitem{Perc05}
W. J. Percival, Astronomy \& Astrophysics, {\bf 443}, 819, (2005)

\bibitem{Mul05}
C. R. Mullis, P. Rosati, G. Lamer, H. B$\ddot {\rm o}$ehringer,
P. Schuecker \&, R. Fassbender, 
Mon. Not. Roy. Astron. Soc., {\bf 623}, L85, (2005);
S. A. Stanford, et al., Astrophys. J., {\bf 646}, L13, (2006)

%

%
\end{thebibliography}
\end{document}